\def\a{\alpha}\def\b{\beta}\def\d{\delta}\def\e{\epsilon}
\def\f{\phi}\def\g{\gamma}\def\h{\theta}
\def\k{\kappa}\def\l{\lambda}\def\m{\mu}\def\n{\nu}\def\o{\omega}\def\q{\psi}\def\r{\rho}\def\s{\sigma}
\def\y{\eta}

\def\coo{coordinates }\def\cor{commutation relations }
\def\des{de Sitter }\def\schr{Schr\"odinger }

\def\mn{{\mu\nu}}\def\ij{{ij}}\def\ha{{1\over 2}}
\def\({\left(}\def\){\right)}\def\[{\left[}\def\]{\right]}

\def\section#1{\bigskip\noindent{\bf#1}\smallskip}

\def\subsection#1{\smallskip\noindent{\it#1}\smallskip}

\def\PL#1{Phys.\ Lett.\ {\bf#1}}

\def\PR#1{Phys.\ Rev.\ {\bf#1}}\def\CQG#1{Class.\ Quantum Grav.\ {\bf#1}}

\def\JMP#1{J.\ Math.\ Phys.\ {\bf#1}}

\def\JoP#1{J.\ Phys.\ {\bf#1}}

\def\EPJ#1{Eur.\ Phys.\ J.\ {\bf#1}}

\def\arx#1{{\tt arXiv:#1}}

\def\ref#1{\medskip\everypar={\hangindent 2\parindent}#1}
\def\beginref{\begingroup
\bigskip
\centerline{\bf References}
\nobreak\noindent}
\def\endref{\par\endgroup}

\def\rom{{\rm m}}
\def\hx{{\hat x}}\def\hp{{\hat p}}\def\hM{{\hat M}}\def\hN{{\hat N}}
\def\cC{{\cal C}}\def\ket{\!>}\def\bra{<\!}\def\ad{a^\dagger}
\def\ulm{{u_{nl\rom}}}\def\eig{|\y_{-1},\y_0,\y_1\ket}
{\nopagenumbers
\line{}
\vskip40pt
\centerline{\bf Reduced Yang model and noncommutative geometry of curved spacetime}
\vskip50pt
\centerline{{\bf S. Meljanac}\footnote{$^\dagger$}{e-mail: meljanac@irb.hr}}
\vskip5pt
\centerline {Rudjer Bo\v skovi\'c Institute, Theoretical Physics Division}
\centerline{Bljeni\v cka c.~54, 10002 Zagreb, Croatia}
\vskip10pt
\centerline{{\bf S. Mignemi}\footnote{$^\ast$}{e-mail: smignemi@unica.it}}
\vskip5pt
\centerline {Dipartimento di Matematica, Universit\`a di Cagliari}
\centerline{via Ospedale 72, 09124 Cagliari, Italy}
\smallskip
\centerline{and INFN, Sezione di Cagliari}
\centerline{Cittadella Universitaria, 09042 Monserrato, Italy}

\vskip80pt
\centerline{\bf Abstract}
\medskip
The Yang model describes a noncommutative geometry in a curved spacetime by means of an orthogonal
algebra $o(1,5)$, whose 15 generators are identified with phase space variables and Lorentz
generators together with an additional scalar generator.
In this paper we show that it is possible to define a nonlinear algebra with the same structure,
but with only 14 generators, that better fits in phase space. The fifteenth generator
of the Yang algebra can then be written as a function of the squares of the others.

As a simple application, we also consider the problem of the quantum harmonic oscillator in this theory,
calculating the energy spectrum in the one- and three-dimensional nonrelativistic versions of the model.
\medskip
{\noindent
}

\vfil\eject}

\section{1. Introduction}
Recently, the Yang model, describing noncommutative geometry in curved spacetime, has revived some interest.
This model was proposed in 1947 by C.N.~Yang [1] with the aim of generalizing to curved spaces
the idea of noncommutative geometry introduced by Snyder [2]. Its original formulation
 was based on an orthogonal
algebra $o(1,5)$ that included phase space operators and the generators of Lorentz symmetries.
A drawback of the formalism was that the closure of the algebra imposed the introduction of a further
generator $K$, whose physical interpretation was not clear.
The mathematical interest of this framework lies in the combination of curved spacetime with curved
momentum space. It may also assume an important physical role as a possible low-energy limit of quantum gravity.

Yang's idea was almost forgotten for fifty years, and it revived when Kowalski-Glikman and
Smolin proposed a model based on a nonlinear quadratic algebra, that contained both de Sitter and Snyder algebra
like the Yang model, but avoided the necessity of the introduction of the generator $K$ [3].
This was called Triply Special Relativity (TSR) or alternatively Snyder-de Sitter model [4].

More recently, a series of papers was devoted to the study of generalizations of the Yang model and
to their realization in quantum phase space and in extended phase space (an extension of phase space
that includes tensorial degrees of freedom) [5,6]. Also its general $\k$-deformations were investigated
in depth in [7].
A review with further references is given in [8].

In this letter, we propose a generalization of the Yang model that, in the same spirit of [3], exploiting
nonlinearity avoids the inclusion of $K$ as an independent generator, and can  be considered a nonlinear
realization of the Yang model.

We also give examples of application
to the dynamics of the harmonic oscillator and compare our results with those obtained for the
related  TSR model [9].

\bigskip

As mentioned above, the Yang algebra is generated by noncommuting position and momentum operators,
Lorentz symmetry generators and one additional Lorentz-invariant operator $K$, and
is isomorphic to an orthogonal algebra.
%However, the generator $K$ has no clear physical interpretation.
In this paper, we show that $K$ can be uniquely expressed in
terms of a function of the squares of the other generators, thus obtaining a new
algebra, whose dimension (in four-dimensional spacetime) is 14, instead of 15
as the original Yang algebra, and can be interpreted a deformation of
the Heisenberg-Lorentz algebra. We call this new nonlinear algebra reduced Yang algebra.
It differs from other nonlinear algebras in phase space related to the Yang model, as for example TSR,
since the deformed Heisenberg algebra has a different form in those cases.

More explicitly, the Yang algebra is defined by the \cor between the phase space \coo $\hx_\m$ and $\hp_\m$,
the Lorentz generators $\hM_\mn$ and $K$ given by\footnote{$^1$}{We adopt the following conventions:
Greek indices run from 0 to 3, Latin indices from 1 to 3, and the flat metric is given by $(-1,1,1,1)$.
We use natural units $c=\hbar=1$.}
$$\eqalignno{&[\hx_\m,\hx_\n]=i\b^2\hM_\mn,\qquad[\hp_\m,\hp_\n]=i\a^2\hM_\mn,\qquad[\hx_\m,\hp_\n]=i\y_\mn K,\cr
&[K,\hx_\m]=i\b^2\hp_\m,\qquad[K,\hp_\m]=-i\a^2\hx_\m,\qquad[\hM_\mn,K]=0,\cr
&[\hM_\mn,\hx_\l]=i(\y_{\m\l}\hx_\n-\y_{\n\l}\hx_\m),\qquad[\hM_\mn,\hp_\l]=i(\y_{\m\l}\hp_\n-\y_{\n\l}\hp_\m),\cr
&[\hM_\mn,\hM_{\r\s}]=i\big(\y_{\m\r}\hM_{\n\s}-\y_{\m\s}\hM_{\n\r}-\y_{\n\r}\hM_{\m\s}+\y_{\n\s}\hM_{\m\r}\big),&(1.1)}$$
where $\a^2$ and $\b^2$ are real parameters, that can take also negative values, and $\y_\mn$ the flat metric.
Clearly, its properties depend on the signs of $\a^2$ and $\b^2$. In fact
the Yang algebra is isomorphic either to $o(1,5)$, $o(2,4)$ or $o(3,3)$,
depending on these signs. It follows that it admits a quadratic Casimir operator
$$\cC=\a^2\hx_\m^2+\b^2\hp_\m^2+{\a^2\b^2\over2}\hM_\mn^2+K^2.\eqno(1.2)$$
From (1.1) it is clear that the position operators do not commute, as in Snyder space, and the same
happens for the momenta, as is expected for de Sitter spacetime. In fact, neglecting $K$, in the limit $\a\to 0$ one recovers
the Snyder algebra [2], while for $\b\to 0$ one obtains the \des algebra.
%We interpret the operators $\hx_\m$ and $\hp_\m$ as \coo of the quantum
%phase space, $M_\mn$ as generators of Lorentz transformations, and $K$
%as a further generator necessary to close the algebra.

\section{2. Reduced Yang model}
The main result of this paper is the proof that the original Yang algebra (1.1)
can be uniquely reduced to a nonlinear algebra with the same form, but where the
generator $K$ is not independent, but is a given function of the other generators.
In fact, the following theorem holds:
\bigbreak
{\noindent \bf Theorem:}

The Lorentz-invariant generator $K$ of the algebra (1) can be uniquely
expressed in terms of the Lorentz invariants $\hx^2=\hx_\m\hx^\m$,
$\hp^2=\hp_\m\hp^\m$ and $\hM_\mn^2=\hM_\mn M^\mn$ as
$$K=\sqrt{1-\a^2\hx_\m^2-\b^2\hp_\m^2-{\a^2\b^2\over2}\hM_\mn^2},\eqno(2.1)$$
so that all the relations (1.1) and the Jacobi identities of the algebra are satisfied.
\medskip
{\noindent \bf Proof:}

It is useful to introduce the Lorentz-invariant operator
$z=\a^2\hx_\m^2+\b^2\hp_\m^2+{\a^2\b^2\over2}\hM_\mn^2$, so that $K=\sqrt{1-z}$.
The operator $z$ satisfies
$$[z,\hx_\m]=-i\b^2(\hp_\m K+K\hp_\m),\qquad[z,\hp_\m]=i\a^2(\hx_\m K+K\hx_\m).\eqno(2.2)$$
Now,
$$\eqalign{&[K,\hx_\m]=[\sqrt{1-z},\hx_\m]\cr
&=i\b^2\sum_{n=1}^\infty(-1)^{n-1}{\ha\choose n}\Big(z^{n-1}(\hp_\m K+K\hp_\m)+
z^{n-2}(\hp_\m K+K\hp_\m)z+\dots+(\hp_\m K+K\hp_\m)z^{n-1}\Big)\cr
&=i\b^2\sum_{n=1}^\infty(-1)^{n-1}{\ha\choose n}\sum_{k=0}^{n-1}
z^k(\hp_\m K+K\hp_\m)z^{n-k-1}=i\b^2\sum_{k,l\ge0}^\infty C_{kl}z^k\hp_\m z^l,}\eqno(2.3)$$
where
$$C_{00}=2{\ha\choose 1}{\ha\choose 0}=1,$$
$$C_{10}=C_{01}=-2{\ha\choose 2}{\ha\choose 0}-{\ha\choose 1}{\ha\choose 1}=0,$$
$$C_{20}=C_{11}=C_{02}=2{\ha\choose 3}{\ha\choose 0}+{\ha\choose 2}{\ha\choose 1}+{\ha\choose 1}{\ha\choose 2}=0.$$
By induction it follows in general
$$C_{kl}=(-1)^{k+l}\sum_{m=0}^{k+l+1}{\ha\choose k+l+1-m}{\ha\choose m}.\eqno(2.4)$$
Then, by the Vandermonde identity follows
$$C_{kl}=(-1)^{k+l}{1\choose k+l+1},\eqno(2.5)$$
which implies $C_{00}=1$ and $C_{kl}=0$ if $k+l\ge1$. Hence,
$$[\sqrt{1-z},\hx_\m]=i\b^2\hp_\m.\eqno(2.6)$$
Analogously,
$$[\sqrt{1-z},\hp_\m]=-i\a^2\hx_\m.\eqno(2.7)$$

In one dimension there is no Lorentz generator and
$$K=\sqrt{1-\a^2\hx^2-\b^2\hp^2}.\eqno(2.8)$$
In the limit $\a=0$, $\b=0$, the reduced Yang algebra becomes the ordinary
Heisenberg algebra together with the Lorentz algebra. In the limit $\a=0$, it reduces
to the Snyder algebra with
$$K=\sqrt{1-\b^2\hp^2}.\eqno(2.9)$$
In the limit $\b=0$, it reduces to the dual Snyder algebra with
$$K=\sqrt{1-\a^2\hx^2}.\eqno(2.10)$$

In general, the reduced Yang algebra can be interpreted as a two-parameter
deformed Heisenberg algebra with exact Lorentz algebra. The $o(1,5)$ Casimir operator $\cC$ takes the value 1.

Clearly, the realizations of this algebra on canonical phase space with \coo $x_\m$ and $p_\m$ coincide with those
of the original Yang model, since the commutation relations are the same.
For example, at first order in $\a^2$ and $\b^2$, a possible Hermitean realization with $\hM_\mn=x_\m p_\n-x_\n p_\m$ is
$$\hx_\m=x_\m-{\b^2\over4}\(x_\m p^2+p^2x_\m\)+\dots,\qquad\hp_\m=p_\m-{\a^2\over4}\(p_\m x^2+x^2p_\m\)+\dots,\eqno(2.11)$$
implying $K=1-\ha(\a^2 x^2+\b^2 p^2)+o(\a^4,\b^4)$.

%Note that in the classical limit, with Poisson brackets replacing commutators, the same algebra can be
%obtained with $M_\mn=\hx_\m\hp_\n-\hx_\n\hp_\m$, while in the quantum case this is not

\section{3. One-dimensional oscillator}
In the  following, we describe the applications of the reduced Yang model to simple quantum mechanical models.
To avoid the intricacies of the relativistic theory, we discuss the Euclidean 3-dimensional case, and for definiteness
we consider  positive
$\a^2$ and $\b^2$. Then the relevant relations are simply obtained by replacing Greek indices by Latin ones in the
previous equations and the symmetry algebra reduces to $o(5)$.

We consider the simplest nontrivial example of dynamics, namely the harmonic oscillator, and start from the
one-dimensional case.
We recall that in standard quantum mechanics, the Hamiltonian of the harmonic oscillator of mass $m$ is [10]
$$H_0=\ha\({p^2\over m}+m\o^2q^2\)={\o\over2}(\ad a+a\ad)=\o\(\ad a+\ha\),\eqno(3.1)$$
where we have defined
$$a=\sqrt{m\o\over2}\(x+i{p\over m\o}\),\qquad\ad=\sqrt{m\o\over2}\(x-i{p\over m\o}\),\eqno(3.2)$$
with
$$[a,\ad]=1.\eqno(3.3)$$
Defining $N=\ad a$, with eigenvectors $|n\ket$, one then gets $N|n\ket=n|n\ket$ and
$$H_0|n\ket\ =\o\(n+\ha\)|n\ket,\eqno(3.4)$$
which gives the energy spectrum of the oscillator.

We proceed in a similar way in the reduced Yang case, by using a perturbative approach.
We start from the Hamiltonian
$$H=\ha\({\hp^2\over m}+m\o^2\hx^2\),\eqno(3.5)$$
and define the ladder operators
$$\hat a=\sqrt{m\o\over2}\(\hx+i{\hp\over m\o}\),\qquad\hat\ad=\sqrt{m\o\over2}\(\hx-i{\hp\over m\o}\),\eqno(3.6)$$
which satisfy
$$[\hat a,\hat\ad]=K.\eqno(3.7)$$
We assume that  $\hx=x+o(\a^2,\b^2)$ and $\hp=p+o(\a^2,\b^2)$, and therefore at order 0, $\hat a=a$ and $\hat\ad=\ad$.

On the other hand, from (2.8), at first order in $\a^2$ and $\b^2$,
$$K\approx1-\ha(\a^2\hx^2+\b^2\hp^2)
%\approx1-{A\over2}(a\ad+\ad a){B\over2}(aa+\ad\ad)
\approx1-{A\over2\o}(\hat a\hat\ad+\hat\ad\hat a)-{B\over2\o}(\hat a\hat a+\hat\ad\hat\ad),\eqno(3.8)$$
where
$$A={\a^2+\b^2m^2\o^2\over2m},\qquad B={\a^2-\b^2m^2\o^2\over2m}.\eqno(3.9)$$
At first order, eq.~(3.7) is satisfied by
$$\hat a=a-{A\over4\o}a\ad a-{B\over8\o}\big(a\ad\ad+\ad\ad a\big),\qquad \hat\ad=\ad-{A\over4\o}\ad a\ad-{B\over8\o}\big(aa\ad+\ad aa\big),\eqno(3.10)$$
and then
$$\eqalign{H&={\o\over2}\big(\hat\ad\hat a+\hat a\hat\ad\big)\approx{\o\over2}(\ad a+a\ad)-{A\over8\o}\big(\ad a \ad a+a\ad a\ad\big)
-{B\over16\o}\big(a^3\ad+a\ad aa+\ad a\ad\ad+(\ad)^3a\big)\cr
&=\o\(N+\ha\)-{A\over2}\(N^2+2N+\ha\)-{B\over4}\(a^2N+(\ad)^2\(N+1\)\)=H_0+H_1,}\eqno(3.11)$$
where $H_0$ is the standard quantum mechanical Hamiltonian  of eq.~(3.1).
At first order, the corrections to the spectrum are given by
$$\bra n|H_1|n\ket\ =-{A\over2}\(n^2+n+\ha\),\eqno(3.12)$$
while the terms proportional to $B$ do not contribute.
Then,
$$E_n=\o\(n+\ha\)-{\a^2+\b^2\o^2\over4m}\(n^2+n+\ha\)+o(\a^2,\b^2).\eqno(3.13)$$
This is the same result obtained in [11] for the Yang model, using a different method based on the solution of the
\schr equation. It is also analogous to the results obtained for TSR [9], where however the corrections have opposite sign.
\bigskip

An interesting special limit is obtained for $\o={\a\over\b m}$. This value of $\o$ is out of the physical range, but
seems to be the most natural choice from an algebraic point of view, and permits to solve exactly the Hamilton equations.
From a purely mathematical point of view, it may be interpreted as the Hamiltonian of a free particle in the reduced Yang
model.

In this case, one has
$$H={1-K^2\over2m\b^2}={1\over2m}\(\hp^2+{\a^2\over\b^2}\hx^2\).\eqno(3.13)$$
Thus $H$ and $K$ commute and have common eigenvectors.
Defining the ladder operators as in (3.2), with $\o={\a\over\b m}$. It follows that
$$[K,a]=\a\b\,a,\qquad[K,\ad]=-\a\b\,\ad.\eqno(3.14)$$
Hence, the role of the number operator is taken in this case by
$$\hN=-{K\over\a\b}.\eqno(3.15)$$
In fact, given an eigenvector $|\y\ket$ of $\hN$, such that $\hN|\y\ket=\y|\y\ket$, one has
$$\hN\ad|\y\ket=(\y+1)\ad|\y\ket.\eqno(3.16)$$

Now, assuming the existence of a ground state $|0\ket$ such that $a|0\ket\ =0$, one has
$$\hN|0\ket\ =n_0|0\ket,\eqno(3.17)$$
and defining recursively $|n+1\ket\ =\ad|n\ket$, it is easy to show that
$$\hN|n\ket\ =(n+n_0)|n\ket.\eqno(3.18)$$
To calculate $n_0$, let us consider the equality
$$K^2=1-2\b^2H=1-2\a\b\,\ad a-\a\b K,\eqno(3.19)$$
which, applied to the ground state, gives
$\a^2\b^2n_0^2=1+\a^2\b^2n_0$ and then
$$n_0=\ha+\sqrt{{1\over4}+{1\over\a^2\b^2}}.\eqno(3.20)$$

Hence,
$$E_n=\bra n|{1-\a^2\b^2N^2\over2\b^2m}\,|n\ket\ ={\a\over\b m}\sqrt{1+{\a^2\b^2\over4}}\(n+\ha\)-{\a^2\over2m}\(n^2+n+\ha\).\eqno(3.21)$$
Therefore, at first order, $E_n\approx{\a\over\b m}\(n+\ha\)-{\a^2\over2m}\(n^2+n+\ha\)$,
which coincides with (3.12) for $\o={a\over\b m}$.

%%%%%%%%%%%%%%%%%%%%%%%%%%%%%%%%%%%%%%%%%%%%%%%%%%%%%%%%%%%%%%%%%%%%%%%%%%%%%%%%%%%%%%%%
\section {4. Three-dimensional oscillator}
Let us now consider the Hamiltonian of a generic harmonic oscillator of mass $m$ in three dimensions.
Unfortunately, to discuss it in a perturbative approach, it is necessary to find a realization of the reduced Yang model
on standard or extended phase space [4], which at leading order reduces to the same as the standard Yang model.

Here, we shall consider a simple Hermitian realization on standard phase space, given by [5]
$$\eqalign{\hx_i=x_i-{\b^2\over4}(p^2x_i+&x_ip^2),\qquad\hp_i=p_i-{\a^2\over4}(x^2p_i+p_ix^2),\cr
&M_\ij=x_i p_j-x_j p_i,}\eqno(4.1)$$
that of course  at leading order satisfies the \cor (1.1).

%The tensorial degrees of freedom are not dynamical and may be decoupled choosing $\g=0$.
We choose a Hamiltonian
$$H=\ha\({\hp_i^2\over m}+m\o^2\hx_i^2+{\g\over2}\hM_\ij^2\).\eqno(4.2)$$
The $M_\ij^2$ term is due to the curvature of position and momentum spaces, and $\g$ is assumed to be linear in $\a^2$ and $\b^2$.
In particular, it is known that the correct Hamiltonian in de Sitter space corresponds to $\g=\a^2/m$.
However, in principle one may add to $\g$ terms proportional to $\b^2$ to take
into account the curvature of momentum space as well [9].
We perturb the Hamiltonian around commutative flat space with $\g=0$.
We have at first order
$$H=H_0-{\a^2+\b^2m^2\o^2\over4m}(x_i^2p_j^2+p_i^2x_j^2+3)+{\g\over4} M_\ij^2,\eqno(4.3)$$
with $H_0=\ha({p_i^2\over m}+m\o^2x_i^2)$ and $M_\ij^2=2x_ip_j(x_ip_j-x_jp_i)$.

In this case, it is more convenient to study the \schr equation. In the unperturbed case, it separates in spherical coordinates, with
$$\q(r,\h,\f)=\sum_{l\rom}{\ulm(r)\over r}\,Y_{l\rom}(\h,\f),\eqno(4.4)$$
where $Y_{l\rom}(\h,\f)$ are spherical harmonics  and $\ulm$ satisfies the equation
$${d^2\ulm\over dr^2}-\(m^2\o^2r^2+{l(l+1)\over r^2}\)\ulm=2mE_{nl}\ulm,\eqno(4.5)$$
with eigenvalues $E_{nl}=\o(2n+l+{3\over2})$ and solutions [12]
$$\ulm=C_{nl}\,e^{-m\o r^2/2}\,r^{l+1}\, L_n^{l+1/2}(m\o r^2),\qquad\qquad C_{nl}^2={2(m\o)^{l+1/2}n!\over(n+l+\ha)!},\eqno(4.6)$$
where $L^\a_n$ are Laguerre polynomials.

One can apply perturbation theory to the Hamiltonian (4.3).
Only diagonal terms contribute to first order.
Denoting $|nl\ket\ =u_{nlm}$ and using (4.5), one has
$$\bra nl|x_i^2p_j^2+p_i^2x_j^2|nl\ket\ =\ 2\bra nl|\,2mE_{nl}r^2-m^2\o^2r^4|nl\ket,\eqno(4.7)$$
and then, from the results of [13],
$$\bra nl|x_i^2p_j^2+p_i^2x_j^2|nl\ket\ =2n^2+2nl+3n+l^2+2l-{3\over4}.\eqno(4.8)$$

It follows that at first order
$$E_{nl}=\o\(2n+l+{3\over2}\)-A\(2n^2+2nl+3n+l^2+2l+{3\over4}\)+{\g\over2}\,l(l+1),\eqno(4.9)$$
with $A$ given by (3.9). Defining a new quantum number $\bar n=2n+l$, this can also be written
$$E_{\bar nl}=\o\(\bar n+{3\over2}\)-{\a^2+\b^2m^2\o^2\over4m}\(\bar n^2+3\bar n+{3\over2}\)+{\g-A\over2}\,l(l+1).\eqno(4.10)$$
Notice that the choice $\g=A$ in the Hamiltonian eliminates the dependence on the quantum number $l$. In fact, with this choice
the symmetry is enhanced, since the Hamiltonian is invariant under transformations generated by $\hp_i$ and $\hx_i$. In
particular, this is the natural choice for $\b=0$.

The spectrum can be compared with that of TSR [9]: like in the one-dimensional case the correction have a similar structure,
but opposite sign.

\section{5. Final remarks}
The Yang model is based on a linear 15-parameters algebra that includes phase space operators, Lorentz generators and an additional
scalar operator that rotates positions into momenta.

We have shown that the Yang algebra can be reduced to a 14-parameter nonlinear algebra, since the scalar generator can be written in terms of the
squares of the other operators in such a way that all the \cor are satisfied.

The dynamics of this model of course differs from that of the standard quantum mechanics. As an example, we have considered the
elementary case of a nonrelativistic harmonic oscillator in 1 and 3 dimensions. The energy spectrum can be calculated perturbatively
and contains corrections proportional to the deformation parameters $\a$ and $\b$. In particular, at first order it coincides with
that of the ordinary Yang model [11]. It is likely that this is true also at higher order.

\beginref

\ref [1] C.N. Yang, \PR{72}, 874 (1947).

\ref [2] H.S. Snyder, \PR{71}, 38 (1947).

\ref [3] J. Kowalski-Glikman and L. Smolin, \PR{D70}, 065020 (2004).

\ref [4] S. Mignemi, \CQG{26}, 245020 (2009).

\ref [5] S. Meljanac and S. Mignemi, \PL{B833}, 137289 (2022).

S. Meljanac and S. Mignemi, \JMP{64}, 023505 (2023).

\ref [6]  J. Lukierski, S. Meljanac, S. Mignemi and A. Pachol, \PL{B847}, 138261 (2023).

T. Martini\'c-Bila\'c,  S. Meljanac and S. Mignemi, \JMP{64}, 122302 (2023).

\ref [7] J. Lukierski, S. Meljanac, S. Mignemi, A. Pachol and M. Woronowicz, \PL{B854}, 138729 (2024).

J. Lukierski, S. Meljanac, S. Mignemi, A. Pachol\ and M. Woronowicz, \arx{2410.02339}.

T. Martini\'c-Bila\'c,  S. Meljanac and S. Mignemi, \EPJ{C84}, 846 (2024).

\ref [8] S. Mignemi, Ukr. J. Phys. {\bf 69},  402 (2024).

\ref [9] S. Mignemi,  \CQG{29}, 215019 (2012).

\ref [10] A. Messiah, "Quantum mechanics", Dover 1999.

\ref [11] S. Meljanac and S. Mignemi, \arx{2411.06443}.

\ref [12] I.J. Zucker, Proc. Camb. Phil. Soc. {\bf 60}, 273 (1994).

\ref [13] D.J. Rowe, \JoP{A38}, 10181 (2008).
\endref

\end

Also in this case (4.3) admits an exact solution for $\o=\g={\a\over\b}$.

The Hamiltonian becomes
$$H=...$$
Let
$$b_1={1\over\sqrt2}(a_1-ia_2),\qquad b_{-1}={1\over\sqrt2}(a_1+ia_2),\qquad b_0=a_3\eqno()$$
with $b^\dagger_\a$ the Hermitian conjugates, where $\a=-1,0,1$.
Then,

$$H= $$
and
$$[b_\a,b_\b]=[b^\dagger_\a,b^\dagger_\b]=0,\qquad[b_\a,b^\dagger_\b]=K\d_{\a\b}$$
$$[K,b_\a]=\a\b b_\a,\qquad [K,b^\dagger_\a]=-\a\b b^\dagger_\a$$

As usual, the $b_\a^\dagger$ can be interpreted as creation operators and we can define a basis  of eigenvectors
$$|\y_{-1},\y_0,\y_1\ket$$
such that, defining $\hat N=-{K\over\a\b}$, one has, using (),
$$\hat N\eig =(\y_{-1}+\y_0+\y_1)\eig$$
$$\hat N\,b^\dagger_\a\eig\ =(\y_{-1}+\y_0+\y_1)\,b^\dagger_\a\eig$$
Hence, $\hat N$...

One can then define a ground state such that
$$\hN|0,0,0\ket\ =N_0|0,0,0\ket,\eqno$$
and defining recursively $|n_{-1},n_0+1,n_1\ket\ =b^\dagger_0|n_{-1},n_0,n_1\ket$, and so on it is easy to show that
$$\hN|n_{-1},n_0,n_1\ket\ =(n_{-1}+n_0+n_1+N_0)|n\ket.\eqno(3.18)$$
To calculate $N_0$, let us consider the equality
$$K^2=1-2\b^2H=1-2\sum_\a\a\b \ad_\a a_\a-3\a\b K,\eqno(3.19)$$
which, applied to the ground state, gives
$\a^2\b^2N_0^2=1+3\a^2\b^2N_0$ and then
$$N_0={3\over2}+\sqrt{{9\over4}+{1\over\a^2\b^2}}.\eqno(3.20)$$

In three dimensions, the ladder operators are defined as
$$a_i=\sqrt{\o\over2}\(x_i+i{p_i\over\o}\),\qquad\ad_i=\sqrt{\o\over2}\(x_i-i{p_i\over\o}\),\eqno(3.2)$$
with
$$[a_i,\ad_i]=\d_\ij\eqno(3.3)$$
and defining $N=\sum\ad_i a_i$, with eigenvalues $|n_i\ket$, one gets $N|n_i\ket=n|n_i\ket$ with $n=n_1+n_2+n_3$ and
$$H_0|n_i\ket\ =\o\(n+{3\over2}\)|n_i\ket\eqno(3.4)$$
which gives the spectrum of the oscillator.

One can define the ladder operators for the Yang model as
$$\hat a_i=\sqrt{\o\over2}\(\hx_i+i{\hp_i\over\o}\),\qquad\hat\ad_i=\sqrt{\o\over2}\(\hx_i-i{\hp_i\over\o}\).\eqno(22)$$
Then, from (1.1),
$$[\hat a_i,\hat\ad_j]=K\d_\ij+{A\over2}M_\ij,\qquad[\hat a_i,\hat a_j]=[\hat\ad_i,\hat\ad_j]=-{B\over2}M_\ij .\eqno(23)$$
with $A$ and $B$...
and
$$H=\sum_i(\hat a_i\hat\ad_i+\hat a_i\hat\ad_i)=\sum\hat\ad_i\hat a_i+{3K\over2}+{\g\over4}M^2_\ij$$
We now expand $a_i$ and $\ad_i$ similarly to (3.6), so that () are satisfied. It follows,
$$\hat a_i=a_i-{A\over4}a_k\ad_ka_i-{B\over8}(a_k\ad_k\ad_i+\ad_i\ad_ka_i),\qquad \hat\ad_i=\ad-{A\over4}\ad a\ad-{B\over8}(aa\ad+\ad aa)\eqno(3.9)$$
and then at first order in $\a^2$ and $\b^2$, the diagonal elements of the operators $K$ and $\g M_\ij^2$ are
$$K=1-\sum_i\({A\over2}(a_i\ad_i+\ad_i a_i)+{B\over2}(a_ia_i+\ad_i\ad_i)\),\qquad\a^2M_\ij^2={3\g\over2}$$
and the diagonal terms of the Hamiltonian (24) can be written as
$$H=\sum_i{\o\over2}(\hat\ad_i\hat a_i+\hat a_i\hat\ad_i)+{3\g\over2}\approx{\o\over2}\[\ad_i a_i+a_i\ad_i-{A\over4}(\ad_i a_i \ad_j a_j+a_i\ad_i a_j\ad_j)
+{3\g\over2}\]$$
Then one gets at first order in perturbation theory
$$E_n=\bra n|\o\(n+{3\over2}\)-{\a^2+\b^2\o^2\over4}\(n^2+3n+{9\over2}\)+3\g+o(\a^2,\b^2).\eqno(3.12)$$
In particular, for $\b\to0$, $\g\to\a^2$, one gets the correct result for a oscillator on $S^3$,
$E_n=\o\(n+{3\over2}\)-{\a^2\over4}\(n^2+3n+{3\over2}\)$ [hi], and for $\a\to 0$, $\g\to0$, the Snyder..
$E_n=\o\(n+{3\over2}\)-{\b^2\o^2\over4}\(n^2+3n+{9\over2}\)$ [minic].
We also notice that choosing $\g=\a^2+\b^2\o^2$ as in [Mi] one gats a simple formula, analogous to.. SDS except for the sign of the corrections.
\end

As in one dimension, one can now obtain the spectrum of $H$ perturbatively in the parameters $\a$ and $\b$, defining
$$\hat a_i=a_i-{A\over4}a_i\ad a_i-{B\over8}(a_i\ad_i\ad_i+\ad_i\ad_i a_i),\qquad \hat\ad_i=\ad_i-{A\over4}\ad a_i\ad_i-{B\over8}(a_ia_i\ad_i+\ad_i a_ia_i)\eqno(3.9)$$

around the standard harmonic oscillator with Hamiltonian
$H_0=\o\(\sum_i\ad_ia_i+{3\over2}\)$.

We recall that the 3-dimensional isotropic oscillator is separable in cartesian coordinates, with eigenvectors $|\{n_i\}\ket=|n_1\ket|n_2\ket|n_3\ket$.
Defining eigenvectors of $H_0$, $|\{n_i\}\ket|l,m\ket$ as direct product of the usual oscillator eigenstates with angular momentum ones,
%$$a^\dagger_i\,|n>|l,m>\ =\sqrt{n_i+1}\ |n_i+1>|l,m>,$$
%$$J_k^2\,|\{n_i\}\ket|l,m\ket\ =l(l+1)\,|\{n_i\}\ket|l,m\ket,\qquad J_+|\{n_i\}\ket|l,m\ket\ =m\,|\{n_i\}\ket|l,m\ket,\eqno(20)$$
at first order the terms proportional to $a_i^2$ and $(a_i^\dagger)^2$ in (27) do not contribute to the perturbative expansion,
and it is easy to check that
$$E_{nlm}=\o\(n_1+n_2+n_3+{3\over2}\)-{3\over4}(\a+\b\o^2)\(n_1+n_2+n_3+\ha\)+\ha\(\g-{3\a\b\over2}\)l(l+1).\eqno(28)$$

An interesting special limit case is obtained for $\o^2=\g={\a\over\b}$. This seems to be the most natural choice
from an algebraic point of view, and may be interpreted as... One has, up to a normalization constant,
$$H={1-K^2\over2}={\b\over2}\hp_i^2+{\a\over2}\hx_i^2+{\a\b\over4}\hM_\ij^2\eqno(29)$$
This is invariant under rotations $\hM_\ij$ and in terms of the Casimir operator of $o(5)$ is given by
$H=\ha{\cC\over1-\a\b}$.

The Hamilton equations read
$$\dot\hx_i=[H,\hx_i]=i\b K\hp_i,\qquad\dot\hp_i=[H,\hp_i]=-i\a K\hx_i,\qquad\dot\hM_\ij=[H,\hM_\ij]=0.\eqno(30)$$
Of course $K$ and $\hM_\ij$ are constant of the motion, while the remaining equations are those of an isotropic
harmonic oscillator with $\o=\sqrt{\a\b}\,K$.

A different approach would be to find a realization of the model on standard phase space, using (12).
In this way one would obtain results that at leading order are identical to those of the standard Yang model,
that has been studied in [] in the 1-dimensional case.
%We also may consider a realization of the \cor in terms of [kempf]...
\end

Let us start from the 1-dimensional case, when tensorial variables are absent. The Hamiltonian reduces to
$$H=\ha\,\hp^2+{\o^2\over2}\,\hx^2\eqno(14)$$
and its spectrum of $H$ can be obtained by standard algebraic methods, defining the ladder operators
$$\hat a=\sqrt{\o\over2}\(\hx+i{\hp\over\o}\),\qquad\hat a^\dagger=\sqrt{\o\over2}\(\hx-i{\hp\over\o}\),\eqno(15)$$
Then,
$$H={\o\over2}\(\hat\ad\hat a+\hat a\hat\ad\),\eqno(16)$$
and
$$[\hat a,\hat\ad]=K.\eqno(17)$$

Therefore, at first order in $\a$ and $\b$, using (9) one has
$$[\hat a,\hat\ad]=1-{\a+\b\o^2\over4\o}(\hat a\hat\ad+\hat\ad\hat a)-{\a-\b\o^2\over4\o}(\hat a\hat a+\hat\ad\hat\ad),\eqno(18)$$
and the Hamiltonian (13) can be written as
$$H=\o\[\(1-{\a+\b\o^2\over4\o}\)\(\hat\ad\hat a+\ha\)-{\a-\b\o^2\over8\o}(\hat a\hat a+\hat\ad\hat\ad)\].\eqno(19)$$

One can then obtain the spectrum of $H$ perturbing in the parameters $\a$ and $\b$ the standard harmonic oscillator with Hamiltonian
$H_0=\o\(\ad a+\ha\)$,where
$$a=\sqrt{\o\over2}\(x+i{p\over\o}\),\qquad a^\dagger=\sqrt{\o\over2}\(x-i{p\over\o}\),\eqno(15)$$
satisfy the standard \cor $[a,\ad]=1$.

Defining as usual eigenvectors $|n\ket$ of $H_0$, such that
$$a\,|n>=\sqrt{n}\ |n-1\ket,\qquad\ad\,|n>=\sqrt{n+1}\ |n+1\ket,\eqno(20)$$
at first order the terms proportional to $a_i^2$ and $(a_i^\dagger)^2$ in (19) do not contribute to the perturbative expansion,
and it is easy to check that
$$E_n=\(\o-{\a+\b\o^2\over4}\)\(n+\ha\).\eqno(21)$$
%A different procedure is to adopt the formalism of [kempf] and write nonstandard ladder operators ...

\bigskip
$$H=\sqrt{\a\over\b}\(N+{K\over2}\)\eqno(23)$$
Then, $N$, $K$ and $H$ commute and have common eigenvectors $|n\ket$. Using (3.6) one gets
$$N|n\ket\ =nK|n\ket\eqno(24)$$

Moreover, from (22) and (23),
$$K^2=1-\a\b(2N+K)\eqno(25)$$
%and then
%$$K={-\a\b+\sqrt{\a^2\b^2-8\a\b N}\over2}\eqno(26)$$
Taking into account (24) one can then obtain
$$K|n\ket=\[-\a\b\(n+\ha\)+\sqrt{1+\a^2\b^2\(n+\ha\)^2}\]|n\ket\ \approx1-\a\b\(n+\ha\)+o(\a^2\b^2)\eqno(27)$$
Finally, defining $K_n=\bra|K|\ket$,
$$E_n=\o\(n+{K_n\over2}\)\approx\sqrt{\a\over\b}\[\(n+\ha\)\(1-{\a\b\over2}\)+o(\a^2\b^2)\]\eqno(28)$$

The spectrum of $H$ can be obtained by standard algebraic methods.
One can define the ladder operators
$$a_i=\sqrt{\a\over2}\,\hx_i+i\sqrt{\b\over2}\,\hp_i,\qquad a_i^\dagger=\sqrt{\a\over2}\,\hx_i-i\sqrt{\b\over2}\,\hp_i,\eqno(14)$$
and the new variables
$$J_i=\e_{ijk}M_{jk},\qquad J_\pm=J_2\pm iJ_3,\eqno(15)$$
such that $J_i^2={\a\b\over2}M_\ij^2$.
Then
$$[a_i,a^\dagger_j]=\sqrt{\a\b}\,K\d_\ij.\eqno(16)$$

Then the Hamiltonian becomes
$$H=\sum_i a_i^\dagger a_i+{3\over2}\sqrt{\a\b}\,K+{\a\b\over2}J_i^2.\eqno(17)$$
The contributions of the vector and tensor \dof separate.

Defining eigenvectors of $H$, $|n_1,n_2,n_3>|l,m>$ in the standard way, such that
$$a^\dagger_i\,|n>|l,m>\ =\sqrt{n_i+1}\ |n_i+1>|l,m>,$$
$$J_i^2\,|n>|l,m>\ =l(l+1)\,|n>|l,m>,\qquad J_+|n,l,m>\ =m\,|n>|l,m>,\eqno(18)$$
the spectrum results in
$$E_{nlm}=(n_1+n_2+n_3+{3\over2}\sqrt{\a\b}K)+{\a\b\over3}\,l(l+1).\eqno(19)$$
\bigskip

A more physical Hamiltonian  can be obtained by rescaling the couplings, as
$$H=\ha\,\hp_i^2+{\o^2\over2}\,\hx_i^2+{\g\over4} M_\ij^2$$
This may be interpreted as the Hamiltonian of a generic harmonic oscillator with both vectorial and tensorial degrees
of freedom. The tensorial degrees of freedom are not dynamical and may be decoupled choosing $\g=0$.

One can proceed as before. Now it is convenient to define the ladder operators in the standard way as
$$a_i=\sqrt{\o\over2}\(\hx_i+i{\hp_i\over\o}\),\qquad a_i^\dagger=\sqrt{\o\over2}\(\hx_i-i{\hp_i\over\o}\)$$
and the $J_i$ as before.
Then, at first order,
$$[a_i,a^\dagger_j]=i\(1-{\a+\b\o^2\over4\o}(aa^\dagger+a^\dagger a)-{\a-\b\o^2\over4\o}(aa+a^\dagger a^\dagger)-{J_k^2\over2}\)\d_\ij$$
and the Hamiltonian becomes
$$H=\(\o-{\a+\b\o^2\over2\o}\)\(\sum_ia_i^\dagger a_i+{3\over2}\)-{\a-\b\o^2\over2}\sum_i(a_ia_i+a_i^\dagger a_i^\dagger)+\ha\(\g-{3\over2}\)J_i^2$$
The contributions of the vector and tensor \dof separate.

One can obtain the spectrum perturbatively in the parameters $\a$ and $\b$. At first order, the terms proportional to
$a_i^2$ and $(a_i^\dagger)^2$ do not contribute, and it is easy to check that
$$E_{nlm}=\o\(1-{\a+\b\o^2\over2}\)\(n_1+n_2+n_3+{3\over2}\)+\ha\(\g-{3\over2}\)l(l+1)$$

\end